# Análisis de los parámetros de los acelerogramas registrados en los seísmos de Lorca, de interés para la Ingeniería

Teresa SUSAGNA[(1)], Luís CABAÑAS[(2)], Xavier GOULA[(1)], Juan Manuel ALCALDE[(2)], Myriam BELVAUX[(3)]

(1) Institut Geològic de Catalunya, c/ Balmes 209-211, 08006, Barcelona, tsusagna@igc.cat.
(2) Instituto Geográfico Nacional, General Ibáñez de Íbero, 3, 28003, Madrid.
(3) BRGM, 3, Avenue C. Guillemin, BP 36009, 45060 Cedex 2, Orleans, Francia.

**Resumen**
La serie sísmica ocurrida en Lorca (Murcia) el día 11 de mayo de 2011 ha permitido registrar numerosos acelerogramas en las distintas estaciones de la red acelerométrica del Instituto Geográfico Nacional (IGN). El análisis de parámetros calculados de manera uniforme ha dado lugar a resultados de interés para la ingeniería sísmica, en particular los correspondientes a los registros de la ciudad de Lorca. La intensidad del movimiento se ha puesto de manifiesto especialmente en la componente horizontal del movimiento, perpendicular a la falla de Alhama de Murcia, que originó el terremoto. Tanto el valor del PGA de de 0,37g, como del CAV de 0,27g•s pueden haberse visto compensados por una duración corta del movimiento de unos pocos segundos, dando lugar a una intensidad macrosísmica no superior a VII en Lorca. La contribución de una componente de campo próximo al movimiento total, debido a la propagación de la ruptura hacia la ciudad, queda reflejada tanto en las señales temporales de aceleración velocidad y desplazamiento como en el espectro de respuesta.
**Palabras clave**: Parámetros de los acelerogramas, Intensidad macrosísmica, Movimiento intenso, Campo próximo, Directividad, Espectros EC8.

## Engineering parameters analysis from accelerograms recorded during the Lorca Earthquake of 11[th] May 2011

**Abstract**
Seismic crisis occurred in Lorca (Murcia) on 11[th] May 2011 originated an important number of accelerograms recorded in the IGN stations. The analysis of uniformly computed parameters has produced interesting results for Earthquake Engineering, in particular those recorded in Lorca. Strong ground motion has been specially observed in the horizontal component perpendicular to the Alhama de Murcia fault, at the origin of the earthquake. Values of PGA= 0,37g and CAV= 0.27g•s seems to be compensated by a short duration of the motion producing a macroseismic Intensity not greater than VII in Lorca. The contribution of near field component of ground motion due to the rupture propagation to and under the Lorca town was shown on acceleration, velocity and displacement time series and also on elastic response spectra.

**Sumario:** Introducción. 1. Análisis de los acelerogramas obtenidos. 2. Parámetros de interés para la Ingeniería 2.1 Dependencia en función de la distancia. 2.1.1 Dependencia de PGA. 2.1.2 Dependencia de la Duración de Trifunac (TD). 2.2 Relación de los parámetros con las intensidades macrosísmicas observadas. 3. Registros en la ciudad de Lorca. 3.1 Espectros de respuesta. Efectos de campo próximo. 4. Discusión y conclusiones. Agradecimientos. 5. Bibliografía.

## Introducción

La serie sísmica ocurrida en Lorca (Murcia) desde el día 11 de mayo de 2011 ha permitido registrar numerosos acelerogramas en las distintas estaciones de la red acelerométrica del Instituto Geográfico Nacional (IGN). Se contabilizan un total de 36 registros (de 3componentes), correspondientes a 13 terremotos de la serie, incluidos el terremoto principal, el premonitor y las réplicas más destacadas.

Se presenta el análisis del conjunto de los registros haciendo énfasis en la interpretación de distintos parámetros: PGA, PGV, IA, CAV, TD, PSV( f) de los acelerogramas del seísmo principal en comparación a los valores conocidos previamente a partir de los registros de la base datos del IGN, de la base europea NERIES, y también en relación con las intensidades macrosímicas observadas. Se hace especial énfasis en los registros y parámetros obtenidos en la ciudad de Lorca.

## 1. Análisis de los acelerogramas obtenidos

El primer terremoto (11/05/2011 15:05:13 UTC; Mw=4.5) fue registrado en 5 estaciones de aceleración, a distancias entre 5 y 40 km, aproximadamente, de su epicentro. El terremoto principal (11/05/2011 16:47:25 UTC; Mw=5.1) fue registrado en 17 estaciones situadas a distancias entre 6 y 185 km. La réplica de mayor tamaño (11/05/2011 20:37:45 UTC; Mw=3.9) fue registrada en 4 estaciones a distancias entre 6 y 39 km. El resto de réplicas con registro de aceleración, todas de magnitudes menores de 3.0, sólo fueron registradas en la estación de Lorca (IGN, 2011).

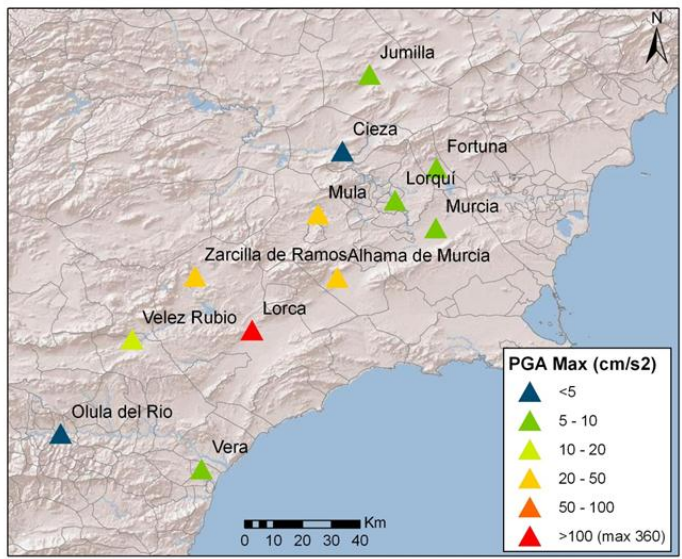

Fig. 1. Estaciones y valores de PGA máximo horizontal del terremoto principal de Mw 5.1.

En la figura 1 se muestran, sobre un mapa regional simplificado, la situación de las estaciones acelerométricas del Instituto Geográfico Nacional (IGN) y los rangos de valores máximos de PGA ("Peak Ground Acceleration") de las componentes horizontales registradas durante el terremoto principal. No están incluidas en dicha figura las estaciones de Jaén y Albolote situadas a más de 180 km y que también registraron el evento principal, con aceleraciones pico pequeñas, del orden de 2 cm/s$^2$. En la Tabla 1 se muestran los valores de aceleración pico sin corregir (PGA_u) para cada dirección o componente del movimiento, en las estaciones que registraron los tres terremotos del día 11 de Mayo, antes citados. La tabla está ordenada por distancias epicentrales de las estaciones.

Tabla 1 Valores pico de aceleración (PGA), registrados en los terremotos más importantes, ocurridos el día 11 de Mayo de 2011.

|  | Dist. Epi (km) | PGA_u (cm/s$^2$) | | |
|---|---|---|---|---|
|  |  | Comp EW | Comp NS | Comp Z |
| **11/05/2011 15:05 Mw=4.5 I=VI** | | | | |
| LORCA(*) | 5 | 128.2 | 270.7 | 75.2 |
| ZARCILLA DE RAMOS | 21 | 10.2 | 6.5 | 8.3 |
| ALHAMA DE MURCIA-02 | 27 | 11.8 | 10.3 | 7.3 |
| VELEZ-RUBIO | 34 | 2.5 | 3.2 | 2.0 |
| MULA | 40 | 7.9 | 6.4 | 5.6 |
| **11/05/2011 16:47 Mw=5.1 I=VII** | | | | |
| LORCA(*) | 6 | 150.8 | 360.3 | 115.1 |
| ZARCILLA DE RAMOS | 21 | 32.1 | 25.4 | 26.2 |
| ALHAMA DE MURCIA-02 | 26 | 44.2 | 41.1 | 23.6 |
| ALHAMA DE MURCIA-01 | 27 | 7.7 | 9.8 | 9.2 |
| VELEZ-RUBIO | 35 | 9.3 | 10.7 | 5.9 |
| MULA | 39 | 41.6 | 35.5 | 20.3 |
| LORQUÍ | 55 | 8.1 | 8.1 | 4.1 |
| VERA | 56 | 7.1 | 5.9 | 4.8 |
| MURCIA | 57 | 8.4 | 7.2 | 3.5 |
| CIEZA | 61 | 2.8 | 2.4 | 1.4 |
| OLULA DEL RIO | 68 | 4.7 | 2.5 | 1.8 |
| FORTUNA | 70 | 7.4 | 6.4 | 3.5 |
| JUMILLA | 89 | 5.4 | 4.1 | 4.4 |
| GUARDAMAR DEL SEGURA | 99 | 1.4 | 2.1 | 0.8 |
| ELDA | 114 | 3.0 | 1.8 | 1.1 |
| ALBOLOTE | 182 | 1.4 | 2.3 | 0.8 |
| JAEN | 185 | 2.8 | 2.1 | 1.3 |
| **11/05/2011 20:37 Mw=3.9 I=IV** | | | | |
| LORCA(*) | 6 | 25.9 | 63.7 | 19.7 |
| ALHAMA DE MURCIA-02 | 26 | 5.9 | 5.7 | 3.3 |
| ZARCILLA DE RAMOS | 21 | 4.3 | 3.5 | 4 |
| MULA | 39 | 5.9 | 5 | 2.7 |

(*) Componentes horizontales tienen orientación E30N y N30W, en lugar de EW y NS

Los valores máximos de PGA registrados en Lorca en los dos primeros terremotos (270 y 360 cm/s$^2$ respectivamente), son los más grandes, jamás antes registrados (instrumentalmente), en la península Ibérica, según consta en las bases de datos del IGN.

Estos altos valores de aceleración pico son, principalmente, consecuencia de la proximidad del terremoto, cuyo epicentro se sitúa a unos 6 km y de la somera profundidad del foco. Después esos valores decaen con la distancia de acuerdo a lo predicho, en general, por los modelos de atenuación del movimiento (ver apartado2.1.1.).

Los acelerógrafos digitales actuales tienen una respuesta plana en un amplio rango de frecuencias. No obstante, los registros de aceleración requieren un cierto proceso, con el fin de eliminar el ruido de baja frecuencia relacionado con la operación del instrumento (ajuste de línea base y filtrado paso-altas). En alta frecuencia se considera que la señal es válida hasta la frecuencia de Nyquist (50 o 100 Hz), aunque en algunos casos esta máxima frecuencia de corte puede rebajarse si el registro presenta ruido en esta zona. En definitiva, en estos procesos (que pueden admitir diversas variantes) se trata de recuperar la señal correspondiente al movimiento del suelo, dentro de una banda de frecuencias adecuada para ingeniería sísmica.

A modo de ejemplo se muestra la componente N30ºO de la estación de Lorca del sismo principal, sólo corregido de línea de base. También se muestran los espectros de respuesta correspondientes para distintos valores de amortiguamiento. En ellos se indica el valor de la frecuencia de corte seleccionado para un filtrado paso altas manual a 0,06Hz (figura 2).

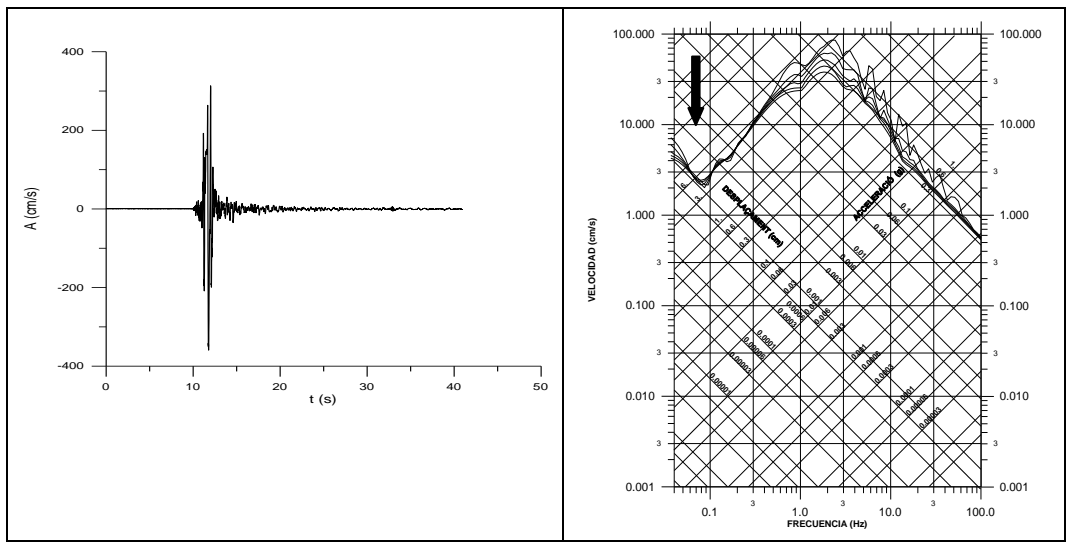

Fig. 2. Acelerograma corregido de línea de base y espectros de respuesta para distintos amortiguamientos asociados, correspondientes a la componente N30ºO de la estación de Lorca del sismo principal. Se indica con una flecha la frecuencia de corte para el filtrado pasa –altas a 0,06Hz.

Como alternativa al proceso manual cuando se necesita procesar un gran número de datos puede utilizarse el programa PARAMACCV9 (Marsal et al., 2008; Roca et al., 2011) desarrollado dentro del proyecto NERIES (2006). En este proyecto se diseñó un sistema distribuido para el intercambio de datos acelerométricos para distintas redes que actúan dentro de la región Euro-Mediterránea. En el marco del proyecto se elaboró un programa estandarizado, con un filtro pasa altas, Butterworth de 2 polos acausal. Se escogió una frecuencia de corte fija para todos los registros de 0.1 Hz.

El procesado de todos los acelerogramas ha sido realizado con este programa. Los registros de la estación de Lorca se han tratado de forma particularizada como se verá en el apartado 3

## 2. Parámetros de interés para la Ingeniería

Para aplicaciones en Ingeniería se han identificado un número importante de parámetros para caracterizar el movimiento del suelo, relacionados a diferentes aspectos del comportamiento estructural de las edificaciones. El programa PARAMACC permite el cálculo de los siguientes parámetros:

- PGA (cm/s$^2$); el más común de los parámetros usados en ingeniería sísmica, usualmente utilizado para escalar las formas espectrales normativas para período nulo,
- PGV (cm/s); una medida alternativa, que refleja el daño en conducciones, contenido de los edificios y elementos no estructurales,
- AI, Intensidad de Arias, (cm/s), una función específica relacionada con el contenido energético (Arias, 1970), a menudo usada como un indicador de la Intensidad Macrosísmica (Cabañas et al., 1997),

$$AI = \frac{\pi}{2 \cdot g} \int_0^\infty a^2(t) dt$$

- TD: Duración de Trifunac o duración significativa (Trifunac y Brady, 1975). Es el intervalo, en segundos, entre el 5% y el 95% de la función de Husid (Husid, 1973), en el que se concentra la mayor parte de la energía,

$$TD(t) = \int_{t\_5\%}^{t\_95\%} a^2(t) dt$$

- CAV: (*Cumulative Absolute Velocity*). Velocidad absoluta acumulada (EPRI, 1988) en cm/s. Esta función está relacionada con el daño estructural al incluir además de la amplitud del movimiento los efectos acumulativos de su duración. (Kramer, 1996; Cabañas et al., 1997),

$$CAV = \int_0^\infty |a(t)| dt$$

- HI, Intensidad de Housner en cm (Housner, 1952), o intensidad de espectro de respuesta. Usada también como medida del daño potencial del acelerograma en estructuras tipo,

$$I_{Housner}(\xi) = \int_{0,1}^{2,5} PSV(\xi, T) dT, \text{ con } \xi = 5\%$$

- PSV(f) Pseudoespectro de respuesta de velocidad para el 5% de amortiguamiento para 28 frecuencias logarítmicamente espaciadas desde 0.15 Hz hasta 39 Hz (desde 0.5 Hz si PGA<0.01g o PGV < 1cm/s).

En la Tabla 2 se presentan los valores calculados de los parámetros siguientes: PGA, AI, TD, CAV, PGV y HI, junto a la distancia epicentral de la estación para el sismo principal y para estaciones situadas a distancias epicentrales menores de 100km.

Tabla 2. Valores de los parámetros PGA, AI, TD, CAV, PGV y HI para el conjunto de registros obtenidos durante el sismo principal y para estaciones situadas a menos de 100km del epicentro.

| Estación | Comp. | dist (km) | PGA (cm/s$^2$) | AI (cm/s) | TD (s) | CAV (cm/s) | PGV (cm/s) | HI (cm) |
|---|---|---|---|---|---|---|---|---|
| LORCA | E | 6 | 151.5 | 11.03 | 3.4 | 163.2 | 14.7 | 32.1 |
|  | N |  | 359.7 | 52.66 | 1.0 | 266.8 | 35.7 | 70.9 |
|  | Z |  | 114.0 | 4.69 | 3.2 | 106.3 | 7.2 | 23.1 |
| ZARCILLA DE RAMOS | E | 21 | 32.0 | 1.47 | 11.8 | 105.2 | 2.1 | 7.5 |
|  | N |  | 25.4 | 1.14 | 11.7 | 92.5 | 2.2 | 9.0 |
|  | Z |  | 26.2 | 0.81 | 11.3 | 75.6 | 1.3 | 5.5 |
| ALHAMA DE MURCIA | E | 26 | 44.1 | 1.31 | 3.6 | 55.7 | 2.1 | 3.0 |
|  | N |  | 41.0 | 1.27 | 3.9 | 55.6 | 1.3 | 3.2 |
|  | Z |  | 23.5 | 0.55 | 6.5 | 42.0 | 0.8 | 1.5 |
| VELEZ RUBIO | E | 35 | 9.3 | 0.14 | 14.8 | 35.1 | 0.6 | 1.8 |
|  | N |  | 10.7 | 0.12 | 18.2 | 32.7 | 0.5 | 2.1 |
|  | Z |  | 5.9 | 0.05 | 16.5 | 22.5 | 0.4 | 1.5 |
| MULA | E | 39 | 41.6 | 1.60 | 9.8 | 104.8 | 1.4 | 3.3 |
|  | N |  | 35.5 | 1.41 | 12.4 | 99.0 | 1.5 | 3.7 |
|  | Z |  | 20.2 | 0.62 | 15.0 | 68.3 | 0.9 | 2.8 |
| VERA | E | 56 | 7.1 | 0.08 | 15.4 | 23.7 | 0.4 | 1.9 |
|  | N |  | 5.8 | 0.06 | 15.8 | 21.9 | 0.4 | 1.9 |
|  | Z |  | 4.5 | 0.04 | 18.6 | 19.0 | 0.3 | 1.0 |
| LORQUI | E | 55 | 8.2 | 0.10 | 22.6 | 34.9 | 0.3 | 1.2 |
|  | N |  | 8.1 | 0.09 | 20.7 | 33.3 | 0.3 | 1.1 |
|  | Z |  | 4.1 | 0.04 | 28.9 | 24.6 | 0.2 | 0.7 |
| MURCIA | E | 57 | 8.4 | 0.07 | 16.2 | 25.4 | 0.4 | 1.0 |
|  | N |  | 7.2 | 0.08 | 14.9 | 26.6 | 0.4 | 1.0 |
|  | Z |  | 3.5 | 0.02 | 27.8 | 16.6 | 0.1 | 0.4 |
| CIEZA | E | 61 | 2.8 | 0.01 | 9.0 | 7.5 | 0.2 | 0.5 |
|  | N |  | 2.4 | 0.01 | 12.3 | 7.8 | 0.1 | 0.4 |
|  | Z |  | 1.4 | 0.00 | 13.9 | 5.4 | 0.1 | 0.3 |
| OLULA DEL RIO | E | 68 | 4.7 | 0.02 | 15.1 | 12.8 | 0.3 | 1.0 |
|  | N |  | 2.5 | 0.01 | 16.2 | 9.4 | 0.3 | 1.0 |
|  | Z |  | 1.8 | 0.01 | 16.8 | 7.5 | 0.2 | 0.9 |
| FORTUNA | E | 70 | 7.5 | 0.10 | 13.5 | 26.7 | 0.3 | 1.1 |
|  | N |  | 6.4 | 0.06 | 17.5 | 23.1 | 0.3 | 0.8 |
|  | Z |  | 3.5 | 0.02 | 18.8 | 14.3 | 0.2 | 0.5 |
| JUMILLA | E | 89 | 5.4 | 0.05 | 15.2 | 20.2 | 0.4 | 1.3 |
|  | N |  | 4.1 | 0.05 | 14.5 | 19.5 | 0.3 | 1.3 |
|  | Z |  | 4.4 | 0.04 | 16.4 | 17.3 | 0.3 | 1.0 |

En la Figura 3 a se ponen en relación los valores de AI y PGA calculados del conjunto de registros del sismo principal. Se muestran junto a la recta de regresión obtenida para el conjunto de registros de la base de datos de acelerogramas europeos NERIES, (Oliveira et al, 2012).

$\log(AI) = 1.83 \cdot \log(PGA) - 2.99 \pm \varepsilon$,  $\rho = 0.98$,  $\sigma_\varepsilon = 0.33$

La correlación parece igualmente válida para los datos del sismo de Lorca, si bien están un poco por encima de la curva media para valores menores de PGA. En la Figura 3b se ha procedido de igual manera con los valores de CAV y de PGA. Se observa también una buena coherencia como en el caso anterior y parecidas desviaciones para valores más pequeños de PGA, respecto a la correlación media de los datos europeos (Oliveira et al., 2012)

$$\log(CAV) = 0.82 \cdot \log(PGA) + 0.48 \pm \varepsilon, \ \rho=0.93, \ \sigma_\varepsilon = 0.28$$

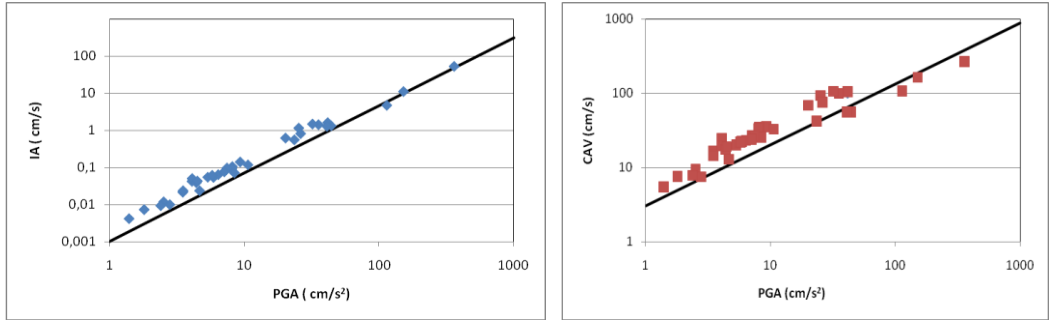

Fig. 3a AI vs PGA de todas las estaciones (componentes horizontales), junto a la recta de regresión obtenida de la base de datos europea NERIES (Oliveira et al., 2012); 3b CAV vs PGA de todas las estaciones (componentes horizontales), junto a la recta de regresión obtenida de la base de datos europea NERIES (Oliveira et al., 2012)

### 2.1 Dependencia en función de la distancia

Se analiza la dependencia de los parámetros PGA y Duración de Trifunac de las componentes horizontales de los acelerogramas registrados durante el sismo principal, en función de la distancia epicentral en comparación con el conjunto de valores correspondientes a los registros existentes en la base de datos del IGN para sismos de Magnitudes comprendidas entre 5 y 5.5.

### 2.1.1 Dependencia de PGA

En la figura 4 se muestra la dependencia para PGA. Se puede observar que para distancias superiores a 20km los valores registrados son parecidos a los valores anteriormente registrados en la red del IGN para sismos de magnitud similar. Los puntos correspondientes a los registros de Lorca que se han representado a una distancia epicentral de 5 km, pero que de hecho podrían representarse a distancias inferiores si tenemos en cuenta la dimensión de la ruptura, muestran valores por encima de los 100cm/s$^2$. No existían hasta el momento registros a distancias tan cortas en la base de datos del IGN.

Si observamos el comportamiento de distintas ecuaciones predictivas del movimiento (GMPE) ajustadas a valores de PGA registrados en distintas zonas, nos damos cuenta que todas ellas son semejantes para distancias comprendidas entre 20km y 100km y concuerdan con los valores observados. Sin embargo hay una gran dispersión para distancias inferiores a 10km, superior a un factor 10 (entre 0,1g y 1 g), debido en gran parte a la profundidad del foco, como reflejan las curvas de Tapia et al. (2007) para profundidades de 0km y 10 km. Los valores registrados en Lorca se acercan más a la curva de profundidad 0km. Observamos también que la curva propuesta por Akkar and Bommer, 2010 se sitúa por encima de las demás y en la parte superior de los datos observados. Cabe señalar que esta última relación se ajustó a datos de sismos de fuerte magnitud contrariamente a las demás relaciones ajustadas en España con datos de magnitudes inferiores a 6.

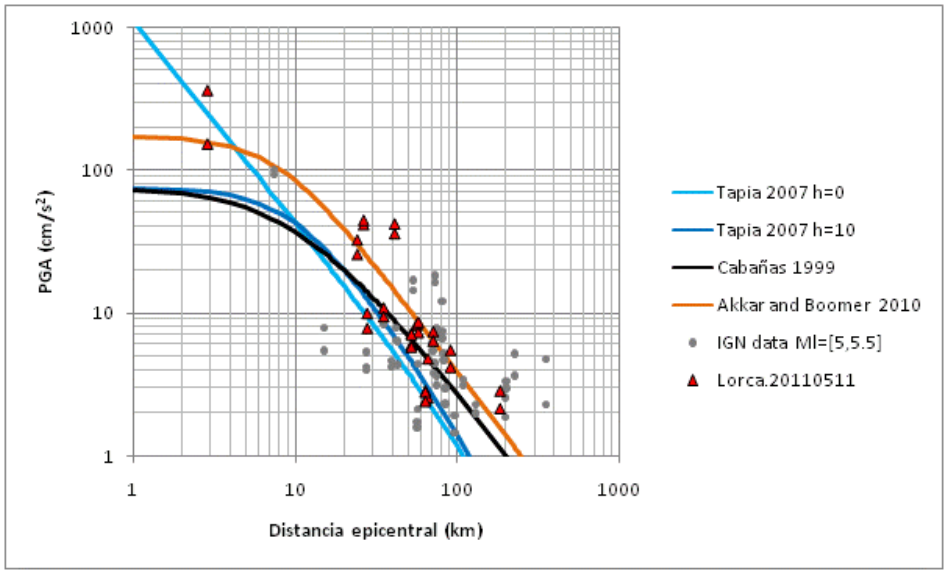

Fig. 4.PGA versus distancia para el conjunto de acelerogramas registrados por la red del IGN para sismos de M5-5.5 (círculos). Los triángulos representan los valores obtenidos para el sismo de Lorca de M5.1. Se representan curvas predictivas del movimiento del suelo (GMPE), según Tapia et al. (2007), Cabañas et al. (1999) y Akkar and Bommer (2010).

### 2.1.2 Dependencia de la Duración de Trifunac (TD)

En la Figura 5 se muestra la dependencia de TD en función de la distancia epicentral. Por definición, este parámetro refleja la duración del movimiento que concentra la mayor parte de energía. Se pone de manifiesto en la figura, una tendencia (obvia) de crecimiento en función de la distancia. Para distancias intermedias, de 20 a 100km, la energía, para sismos de M5-5.5, se concentra en 10-20s, mientras que para los pocos registros existentes para distancias menores a 10km la duración es sólo de unos pocos segundos.

Esta constatación puede tener su importancia en los daños observados, reflejados en la intensidad macrosísmica asignada a la ciudad de Lorca no superior a VII (EMS'98). La corta duración del movimiento puede haber compensado su gran amplitud dando lugar a una aceleración "efectiva" (asociada al daño), inferior al valor máximo observado de 0,37 g.

Otra implicación de esta constatación es que las señales temporales utilizadas comúnmente en cálculo de estructuras tienen duraciones de 20-30 s, mucho más largas que las registradas en Lorca. La modelización de los daños causados en estructuras en Lorca ha de tener en cuenta la corta duración del acelerograma. Sería probablemente, la manera de verificar si la fuerte amplitud del movimiento que entra en la estructura se ha visto compensada por su corta duración habida cuenta de la ductilidad de la estructura.

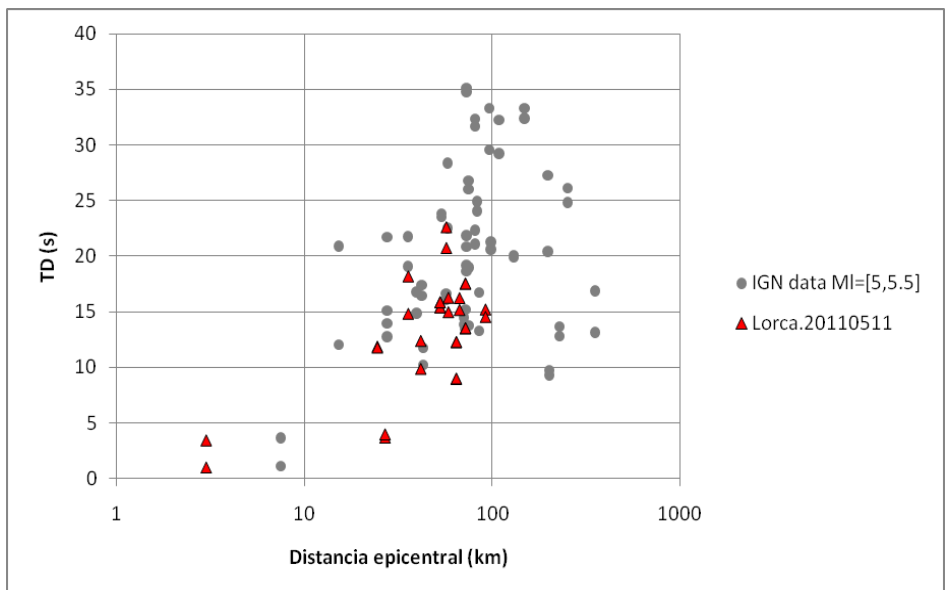

Fig. 5. TD versus distancia para el conjunto de acelerogramas registrados por la red del IGN para sismos de M5-5.5 (círculos). Los triángulos representan los valores obtenidos para el sismo de Lorca de M5.1

## 2.2 Relación de los parámetros con las intensidades macrosísmicas observadas

El sismo principal fue ampliamente percibido por la población en una gran zona en torno al epicentro. Las determinaciones de intensidades macrosísmicas fueron realizadas por el IGN a partir de los cuestionarios recibidos telemáticamente e interpretados automáticamente, especialmente para valores de intensidad de hasta el grado V. Para las zonas más próximas al epicentro y con daños en los edificios estas informaciones fueron completadas por la visita de técnicos especialistas a las poblaciones afectadas. Los mapas de intensidades pueden consultarse en la web del IGN ( www.ign.es).

En la figura 6 se muestran los valores estimados de la intensidad macrosísmica en aquellas poblaciones dónde también se dispone de registros acelerométricos.
Estas informaciones contribuyen de manera importante a estimar correlaciones con
parámetros de interés en ingeniería, en particular para terremotos de esta magnitud, para
los que se dispone de poca información en España

Así pues se muestran en las figuras 7, 8a y 8b los valores de intensidad respecto a los de los parámetros PGA, AI y CAV calculados en las componentes horizontales de los acelerogramas registrados.

En la Figura 7 se presentan los datos obtenidos en el sismo principal, conjuntamente con los datos disponibles en la base de datos del IGN. La primera observación es la presencia de un número importante de datos para intensidades inferiores o iguales a V que presentan una gran dispersión. Las observaciones de Lorca para estas intensidades entran dentro de los valores ya conocidos. Para intensidades superiores hay que señalar la poca información disponible. Los datos de la ciudad de Lorca son los únicos disponibles para la intensidad VII.

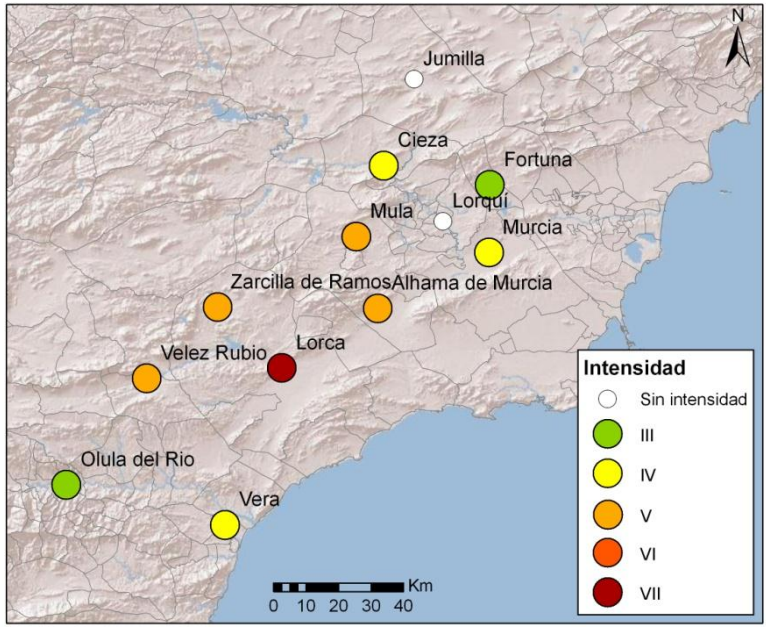

Fig. 6. Intensidades percibidas en los municipios dónde se dispone de acelerogramas registrados (IGN, 2011)

Existen en la literatura muchas tentativas de ajustes de relaciones entre Intensidad y PGA. Nos hemos interesado solamente a unas pocas por distintas razones:
- la relación utilizada en España en las distintas versiones de la Norma de Construcción Sismoresistente (NCSE) y todavía utilizada para realizar el mapa de peligrosidad propuesto en la NCSE02:
    - $\text{Log}_{10}\, a_b = 0{,}301030 * \text{Int} - 3{.}2233$
- Unas relaciones propuestas por Souriau (2006), ajustadas con datos más recientes del sur de Francia para valores de intensidad inferiores o iguales a V y por tanto con estimaciones de intensidad parecidas a las españolas por tratarse de la misma escala. La relación propuesta tiene dependencia de la distancia epicentral, R,
    - $\log_{10}(\text{PGA}) = -1.78(\pm 0.39) + 0.37(\pm 0.09)\text{I} - 0.45(\pm 0.10) \log_{10} R$
- La relación de Wald et al., (1999) ajustada en EEUU y aplicadas para todo el mundo en los ShakeMap publicados por el USGS. Las relaciones están ajustadas principalmente para movimientos intensos (8 terremotos de California entre 1971 y 1994) y por tanto para valores altos de la intensidad. Hay que tener en cuenta la falta de homogeneidad de escalas y también las diferencias de tipologías constructivas entre EEUU y España.
    $I_{MM} = 3.66 * \log_{10}(\text{PGA}) - 1.66$ para $V \leq I_{MM} \leq VIII$

$$I_{MM}=2.20 *\log_{10}(PGA)+1.00 \quad \text{para } V \leq I_{MM}$$

En la Figura 7 se han representado las relaciones NCSE, Wald et al. (1999) y Souriau (2006), para distancias epicentrales de 10 y 100km.

La relación NCSE ajusta bien los datos para intensidades menores o iguales a V, pero asocia valores débiles de PGA para grandes intensidades. Presenta la ventaja de estar acorde con la conocida regla usada en ingeniería, de aumentar 1 grado de intensidad cuando se dobla el valor de la aceleración. Se puede observar que la intensidad VII para esta relación corresponde a un PGA de 0.07g, muy por debajo de los valores observados en Lorca.

Las curvas de Souriau (2006) ajustadas sólo a valores de movimientos débiles presentan una pendiente algo distinta. En este caso, la extrapolación de las curvas que dependen de la distancia, da valores de PGA entre 0,08 y 0,22g para una intensidad de VII.

Finalmente, la curva propuesta en EEUU por Wald et al., (1999) propone un cambio de pendiente a partir de la Intensidad de V, es decir para movimientos con daños asociados. A la intensidad VII se asocia un PGA próximo a =0.22g. Se corresponde al valor medio del PGA observado en las dos componentes horizontales de Lorca.

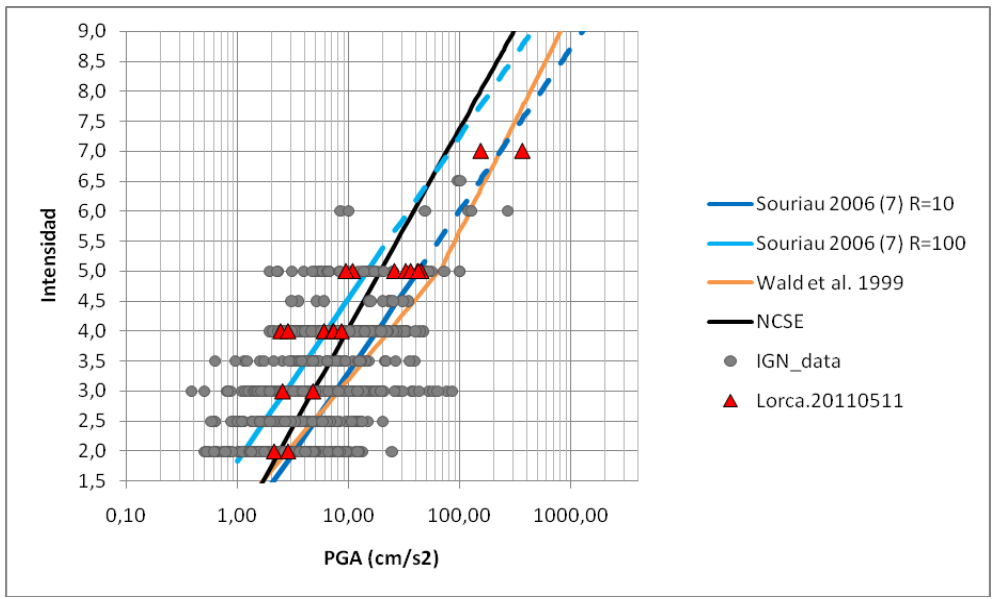

Fig. 7. Intensidades versus valores de PGA (componentes horizontales) observados para el conjunto de sismos existentes en la base de datos del IGN. En rojo, los puntos correspondientes a las observaciones del sismo de Lorca junto con las curvas de NCSE, Souriau (2006) y Wald et al. (1999).

En definitiva las curvas propuestas que mejor ajustan datos de intensidades inferiores a V son la NCSE y Souriau (2006) para distancias medias. La relación de Wald et al. (1999) asocia valores altos de PGA a estas intensidades. Para valores de intensidad mayores que V, las curvas que mejor ajustan a los pocos datos disponibles, son Souriau (2006) para distancias medias y Wald et al. (1999). Para valores muy elevados de intensidad la extrapolación de Souriau (2006) da valores de PGA un poco altos, (0,7g para Intensidad IX). La relación de Wald et al. (1999) da valores parecidos.

En trabajos previos, por ejemplo Cabañas et al., (1997) se han estudiado correlaciones entre los parámetros AI y CAV con la intensidad, con el objetivo de relacionar el daño con parámetros simples deducidos de los acelerogramas.

En la figura 8 a) y b) se muestran respectivamente los valores de los parámetros AI y CAV calculados en las componentes horizontales de los acelerogramas junto con las relaciones propuestas por Cabañas et al., (1997) ajustados a datos de sismos italianos. Hay que señalar que ese trabajo utiliza para el CAV una modificación del parámetro, denominado CAV20 (sólo contribuyen a la integral aquellas ventanas de 1 s de duración que alcancen al menos un valor de aceleración superior a 0, 02 g). Este tipo de filtrado fue propuesto por EPRI (1991), para un umbral de aceleración de 0.025 g, con el fin de eliminar del valor, la contribución de registros de aceleración con largas codas de baja amplitud. El parámetro fue denominado $CAV_{STD}$ (estandarizado) y un valor de 0.16 g·s fue fijado entonces como un umbral de daño en estructuras de ingeniería bien diseñadas.

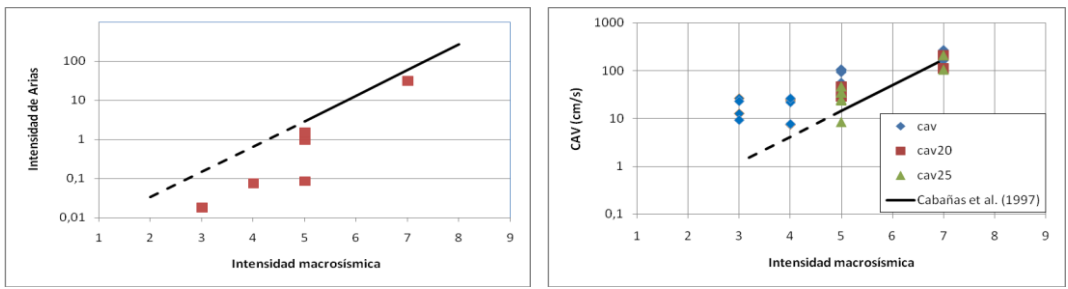

Figura 8 a) Valores de la Intensidad de Arias (AI) versus intensidad calculados para componentes horizontales de los acelerogramas registrados en el sismo de Lorca junto con la relación media establecida por Cabañas et al, 1997; b) Valores del parámetro CAV versus intensidad calculados para componentes horizontales de los acelerogramas registrados en el sismo de Lorca, junto con la relación media establecida por Cabañas et al., 1997

Para la Intensidad de Arias (AI) (figura 8a), los puntos observados en Lorca para valores superiores a 1 cm/s se asemejan a la relación propuesta por Cabañas et al. (1997): ln (AI)= 1.50 $I_L$-6.42, $R^2$ =0,92. Cabe señalar que esta relación es usada en estudios de microzonación sísmica (Figueras et al., 2012; Goded et al., 2012) para obtener incrementos del valor de la Intensidad a partir de registros calculados en suelo en relación con los registros obtenidos en roca. Sólo el coeficiente 1,5 representando la pendiente de la recta de regresión importa para esta estimación. Para valores más bajos de AI, la extrapolación de la recta no parece ajustar tan bien a los datos de Lorca.

Para CAV (figura 8b), los valores calculados de los registros obtenidos en el sismo de Lorca muestran valores muy dispersos según el tipo de filtrado. Estos valores están por encima de la relación propuesta para CAV20 por Cabañas et al. (1997), especialmente para movimientos débiles (intensidad V). Para movimientos más intensos, como el registrado en Lorca de Int=VII, los valores de CAV dependen menos del filtrado y son muy parecidos a los previstos en la dicha relación realizada para sismos italianos. Las diferencias de filtrado son más visibles para movimientos débiles, dónde la contribución de valores inferiores al umbral fijado (0,02g o 0.025g) es más importante.

El valor de 0.16 g•s para $CAV_{STD}$ sigue siendo utilizado en la industria nuclear, (EPRI 2006) como un umbral (conservador) del comienzo del daño y por tanto como umbral de seguridad básica. Los valores calculados en los acelerogramas de la estación de Lorca para las componentes horizontales ofrecen valores de 0,17 g•s y 0,27 g•s para CAV (sin filtrado) y de 0,11 g•s y 0,22 g•s para $CAV_{STD}$. Es la primera vez que en España se detectan estos valores para un sismo de magnitud moderada.

## 3. Registros en la ciudad de Lorca

Los registros obtenidos en el acelerógrafo situado en el sótano de la antigua cárcel de la ciudad de Lorca han permitido analizar el movimiento del suelo en campo cercano a una distancia de unos 6 km del epicentro.

Como se ha visto en la Figura2 el espectro del registro sin filtrar permite asignar el filtro más conveniente en el proceso de las señales. Se ha aplicado un filtro paso-altas a la aceleración antes de realizar las integraciones para obtener la velocidad y el desplazamiento. Para el sismo principal se ha definido la frecuencia de corte en 0.06 Hz.

En la tabla 3 se muestra la estimación de la aceleración, velocidad y desplazamiento de los tres sismos más importantes de la serie. Siempre viéndose un mayor valor en la componente N30ºO

Tabla 3. Valores de aceleración, velocidad y desplazamiento para los tres sismos más importantes de la serie sísmica de Lorca

| sismo | comp. | Acel. [cm/s2] | Vel (cm/s) | Des (cm) |
|---|---|---|---|---|
| Premonitorio | E30ºN | 128.03 | 4.05 | 0.31 |
| 15:05:13 | N30ºO | 270.39 | 12.99 | 0.98 |
| Mw 4.5 | V | 75.03 | 2.31 | 0.24 |
| Principal | E30ºN | 151.18 | 14.36 | 1.26 |
| 16:47:25 | N30ºO | 359.48 | 35.76 | 2.99 |
| Mw 5.1 | V | 114.16 | 7.61 | 1.19 |
| Réplica | E30ºN | 25.89 | 0.81 | 0.04 |
| 20:37:45 | N30ºO | 63.59 | 1.78 | 0.13 |
| Mw 3.9 | V | 19.64 | 0.55 | 0.04 |

En la Figura 9 se presenta las 3 componentes de los registros de aceleración, velocidad y desplazamiento. Son de destacar los valores máximos registrados en la componente horizontal, de azimut N30ºO, casi perpendicular a la dirección de la falla y de la ruptura, debido muy probablemente a que la ruptura sísmica se propagó por debajo del núcleo urbano de Lorca, hasta una distancia muy próxima al emplazamiento.

Tal como se muestra en las Figuras 9 y 10, el desplazamiento máximo observado en la componente horizontal N30ºO es de al menos 3 cm (el filtrado operado en el registro de aceleración puede haber eliminado parte del desplazamiento real). Esencialmente el movimiento observado es un pulso simétrico de una duración de algo más de 1 s. El gran desplazamiento observado en la componente casi perpendicular a la dirección de la ruptura puede ser debido probablemente a un efecto de directividad, causado por la propagación de la ruptura en dirección a la ciudad de Lorca. Se pudo observar el efecto de este desplazamiento dinámico horizontal sobre distintos edificios y maquinaria pesada, como se ve en la figura 11 (IGC, 2011).

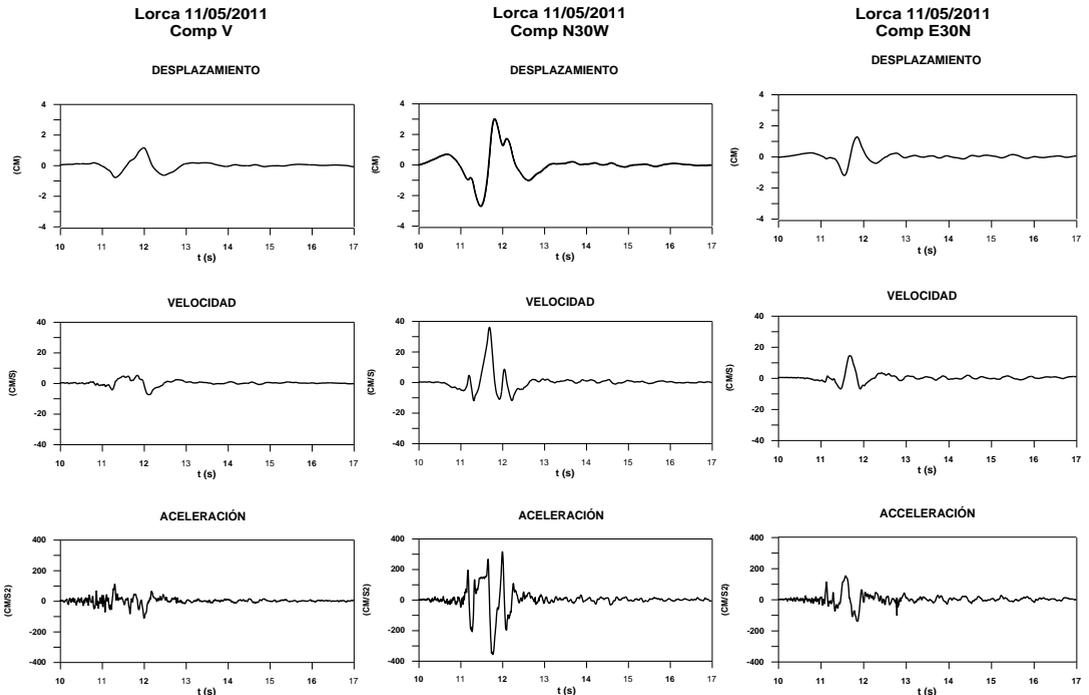

Fig. 9. Registros de aceleración, velocidad y desplazamiento (3 componentes), obtenidos en el acelerógrafo de Lorca.

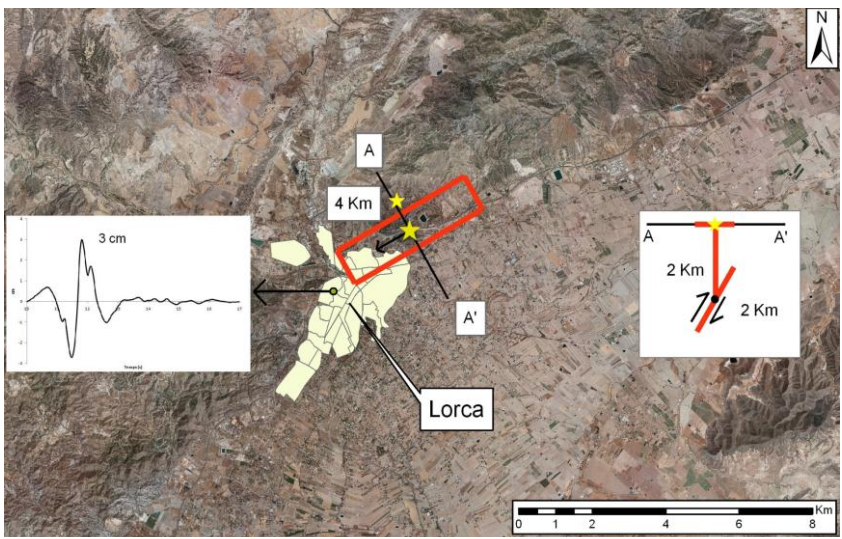

Fig. 10. Esquema de la posición del epicentro y de la ruptura sísmica con las dimensiones estimadas y el registro del desplazamiento correspondiente a la componente horizontal N30ºO (Frontera et al., 2012)

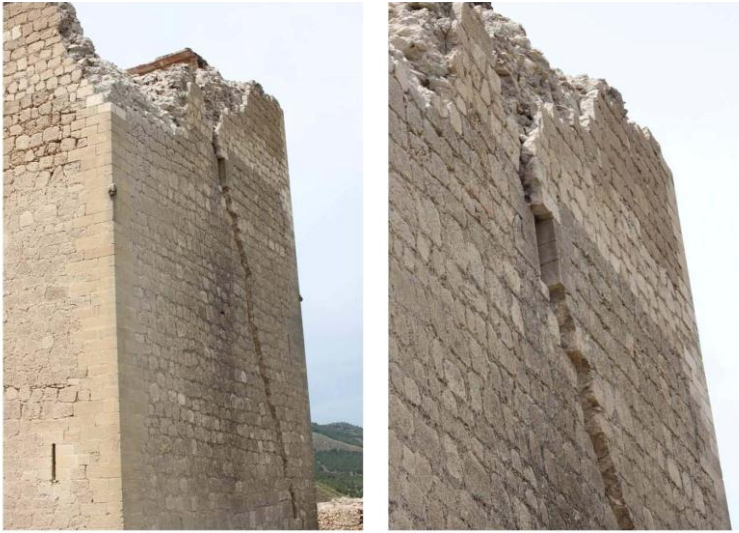

Fig. 11. Paredes del castillo agrietadas en la dirección del movimiento observado, mostrando un desplazamiento de algunos centímetros (IGC, 2011)

El fenómeno de directividad, cuando se manifiesta, afecta al movimiento del suelo especialmente en la dirección horizontal y perpendicular a la falla dónde se origina el terremoto. El segmento de la falla de Alhama que rompió durante el sismo presenta un acimut N60ºE y la ruptura se propagó principalmente de Este a Oeste. Es pues lógico observaren el registro de la estación de Lorca, un movimiento del suelo más intenso en la componente N30ºO que en la componente N60ºE.

Es también interesante de señalar que además del desplazamiento dinámico de unos 3 cm se ha observado un desplazamiento permanente de unos pocos mm en la estación CGPS de Lorca (Frontera et al., 2012). Cabañas et al. (2012) han logrado poner de manifiesto la presencia de desplazamiento de varios mm a partir del acelerograma registrado en Lorca con un proceso adaptado a tal fin.

### 3.1 Espectros de respuesta. Efectos de campo próximo

La aceleración máxima del suelo observada en el registro de la componente N30ºO de la aceleración (Figura 9) ha alcanzado un valor de 0,37g. La mayor parte de la energía ha sido liberada en un intervalo de tiempo de algo más de 1 s en 3 pulsos principales.

Este valor de la aceleración máxima es muy superior al valor de la aceleración básica, $a_b$, previsto en la Norma de Construcción Sismorresistente Española (NCSE-02) para Lorca de 0,12g.

En la Figura 12 se han representado los valores espectrales para un amortiguamiento del 5% para las tres componentes junto con el espectro EC8 de tipo 2 para clase de suelo A (roca), escalado al PGA de la componente N30O.

Cabe destacar las siguientes observaciones:
- la forma espectral del EC8 se ajusta bastante bien al espectro observado excepto que la "meseta" del EC8 presenta una anchura menor,
- presencia de un pulso en las tres componentes registradas, en torno a 0,4 s.

La presencia en el espectro de un pulso a 0,4 s puede ser una consecuencias del fenómeno de directividad, ya señalado previamente en la descripción de la componente N30ºO del desplazamiento si bien en los espectros se observa en todas las componentes. Rueda et al. (2001) han puesto igualmente en evidencia un pulso de directividad usando el método de "wavelets". El período de este tipo de pulso aumenta generalmente con la magnitud, ya que su valor está asociado al tiempo de subida de la dislocación y a las dimensiones de la falla, que aumentan con la magnitud. Para una magnitud igual a 6 el pulso es cercano a 0,8 s. Puede aumentar hasta 4 s para una magnitud de 7.5. En el caso de Lorca (M5.1), la observación de un pulso a un período de 0,4 s es compatible con esos órdenes de magnitud (AFPS, 2011).Por tanto, los espectros observados en Lorca se diferencian de la forma espectral EC8 (2003) tipo2 en el ancho de meseta, probablemente en parte, debido a la contribución del efecto de campo cercano.

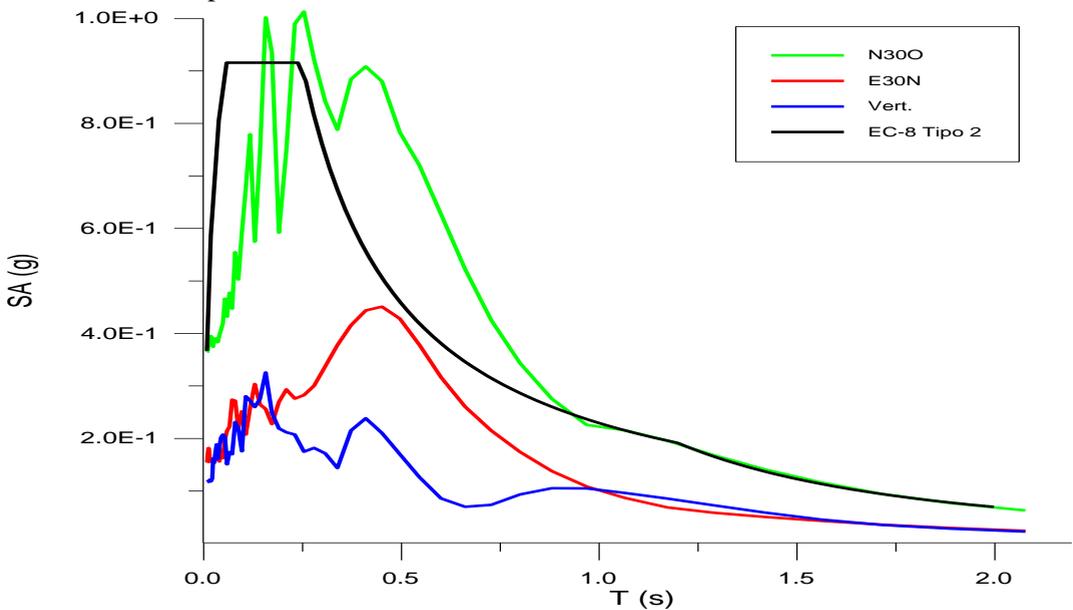

Fig. 12. Espectros de respuesta de las tres componentes de los acelerogramas registrados en Lorca junto con la forma espectral EC8 de tipo 2 para clase de suelo A (roca), escalado al PGA de 0,37g

Otra manera de entender esta diferencia la sugieren Chai et al. (2000). En efecto, el análisis de los acelerogramas registrados en campo cercano durante el sismo de Chi- Chi de 1999 en Taiwan, condujo a los autores a proponer modificaciones a las formas espectrales de la normativa de Taiwan que se concretan en tres factores de amplificación: uno, en altas frecuencias (PGA), otro, para frecuencias intermedias (meseta) y un tercero para períodos largos; también se propone una mayor anchura de la meseta del espectro.

Ello queda reflejado en la figura 13. En el caso de Lorca tal como se ha visto en la Figura 10 las amplificaciones se han observado en períodos cortos (PGA= 0,37g), en períodos intermedios y claramente en la anchura de la "meseta". No se observa en cambio la amplificación para períodos largos, ya que el espectro coincide perfectamente con la forma espectral tipo2 del EC8 entre 1 y 2 s de período (ver figura 12).

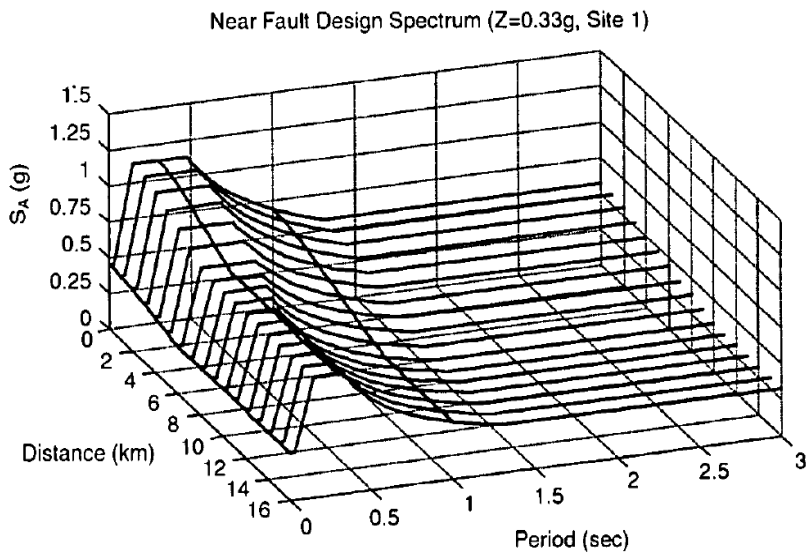

Fig, 13. Espectros de diseño para emplazamientos cercanos a la falla (Chai et al. 2000).

Hay que señalar igualmente que la Normativa de USA, Uniform Building Code, (1997) prevé modificaciones en los espectros de diseño para emplazamientos próximos a fallas activas a través de unos coeficientes de amplificación (*near source factors Na y Nv*) que se aplican a las ordenadas espectrales de periodos cortos e intermedios y que son parecidos a los propuestos en Taiwan. En este caso los coeficientes dependen del tipo de falla activa de que se trate (clases A, B, C) y de la distancia a la misma.

**4. Discusión y conclusiones**
Los acelerogramas obtenidos en la región de Lorca durante los sismos del mes de Mayo de 2011 han permitido aumentar con datos de movimientos intensos las bases de datos existentes en España, escasos hasta la fecha. Se han podido calcular de manera uniforme parámetros de interés para la ingeniería: PGA, AI, CAV HI, TD, PSV (f) con el programa PARAMACC puesto a punto en el IGC (Marsal et al, 2008) en el proyecto europeo NERIES, para la constitución de una base datos distribuida en Europa (Roca et al., 2011). Su análisis ha dado lugar a un número importante de observaciones que han podido compararse a observaciones y resultados previos. Destacamos los siguientes resultados:
- Correlaciones entre AI vs PGA y CAV vs PGA de todos los registros obtenidos en el sismo de Lorca muestran una buena coherencia con correlaciones previas realizadas con el conjunto de datos europeos (Oliveira et al., 2012)

- La dependencia de los parámetros PGA y TD en función de la distancia para los datos del sismo de Lorca muestran una buena concordancia para distancias medias con los valores conocidos previamente de la base de datos del IGN, así como con la comparación de diversas curvas GMPE propuestas para España y para la región euroMediterránea Los registros de la ciudad de Lorca aportan nuevas informaciones para distancias cortas, muy escasos hasta el presente. Para PGA, la relación de Akkar and Bommer (2010) así como Tapia et al. (2007) para profundidades cercanas a 0 km ajustan bien a estas observaciones. Los datos TD muestran que la energía de registros próximos se concentra en unos pocos segundos.
- La compensación de fuertes amplitudes con las duraciones de pocos segundos puede contribuir a mitigar la importancia del daño observado. La corta duración del acelerograma debe tenerse en cuenta en los cálculos de modelización del daño en las estructuras de Lorca.
- Se ha podido comparar las intensidades macrosísmicas estimadas en diversas poblaciones con los valores de los parámetros de los acelerogramas registrados. La representación de los valores de Intensidad vs PGA junto con los disponibles en la base datos del IGN ha permitido observar la gran dispersión de las observaciones para intensidades inferiores a V y destacar la contribución de los datos de Lorca para movimientos intensos. Las curvas actualmente utilizadas en España por las distintas NCSE no parecen adaptadas a movimientos intensos. Se han discutido la adecuación de relaciones propuestas por Souriau (2006) y Wald et al. (1999) para intensidad de VII, así como la dificultad de extrapolarlas para mayores intensidades.
- El análisis de los espectros obtenidos en Lorca han permitido observar la contribución de la directividad de la ruptura en ellos en forma de un pulso cercano a T=0.4s. Este pulso constituye la mayor diferencia del espectro respecto a la forma espectral de tipo2 EC8 para roca. Esta observación, junto con el aumento del PGA, se compara con las observaciones realizadas en Taiwan durante el sismo de Chi-Chi, que contribuyeron a la propuesta de modificación de los espectros normativos en zonas cercanas a fallas activas.

## Agradecimientos

## 4. Bibliografía

AFPS (2011). Le seisme de Lorca (Espagne) du 11 mai 2011. Rapport de mission. Décembre 2011. 152 pp.

Akkar, S. and Bommer, J. (2010). Empirical Equations for the Prediction of PGA ,PG V, and Spectral Accelerations in Europe, the Mediterranean Region, and the Middle East. Seismological Research Letters, vol 81, nº2.

Arias, A. (1970). A measure of earthquake intensity. In: RJ Hansen (ed) Seismic design for nuclear power plants. MIT Press, Cambridge, MA, pp 438–483.

Cabañas, L., Benito, B., Herráiz, M. (1997). An approach to the measurement of the potential structural damage of earthquake ground motion, Earthquake Engineering and Structural Dynamics, 26:79-92.

Cabañas, L., Alcalde, J. M., Carreño, E. and J.B. Bravo (2012). Characteristics of observed strong motion accelerograms from the 2011 Lorca (Spain). Bulletin of Earthquake Engineering. (Submitted)

Chai, J., Loh, C., Chen, C. (2000). Consideration of the near-fault effect on seismic desing code for sites near the Chelungpu fault. Journal of the Chinese Institute of Enginneers, Vol 23, nº4, 447-454.

EC8 Eurocode 8 (2003). Design of Structures for Earthquake Resistance. Comité Européen de Normalisation.

EPRI (1988). A Criterion for Determining Exceedance of the Operating Basis Earthquake, Report No. EPRI NP-5930, Palo Alto, California.

EPRI (1991). Standardization of the Cumulative Absolute Velocity, Report No. EPRI TR-100082-T2, Palo Alto, California.

EPRI (2006). Program on Technology Innovation: Use of Cumulative Absolute Velocity (CAV) in Determining Effects of Small Magnitude Earthquakes on Seismic Hazard Analysis. Technical Report.

Figueras, S., Macau, A., Peix, M., Belvaux, M., Benjumea, B., Gabàs, A., Susagna, T. y Goula, X. (2012). Caracterización de efectos sísmicos locales en la ciudad de Lorca. Física de la Terra, 24. En este mismo volumen

Frontera, T., Concha, A., Blanco, P., Goula, X., Arbiol, R., Pérez, F. (2012). Deformaciones cosísmicas del terremoto de Lorca del 11 de mayo de 2011. Física de la Terra, 24. En este mismo volumen.

Goded, T., Buforn, E., Macau, A. (2012). Site effects evaluation in Malaga city's historical centre (Southern Spain). Bul. Earthquake Eng. DOI 10.1007/s10518-011-9337-4

Housner, G.W. (1952). Spectrum intensities of strong-motion earthquakes. In Proceedings of Symposium on Earthquake and Blast Effects on Structures. Earthquake Engineering Research Institute, Berkeley, CA.

Husid, R. (1973). Terremotos: análisis spectral y características de acelerogramas como base de diseño sísmico. Andres Bello, 447.

IGC (2011). Terremoto de Lorca, España, 11 de mayo de 2011: Informe técnico de inspección de campo. Monografia nº 3, IGC ·86 pp.

IGN (2011). Informe del sismo de Lorca del 11 de Mayo. www.ign.es

Kramer, S.L. (1996). Geotechnical earthquake engineering. Prentice Hall Inc, Upper Saddle River, NJ, 651 pp.

Marsal, A., Susagna, T., Goula, X., Oliveira, C.S. (2008). Implementation of Accelerometric Parameters. Computation and Exchange. NA5-D4 Report, NERIES.

NERIES Project (2006) Network of Research Infrastructures for European Seismology. (http://www.neries-eu.org/). (Accessed 27 January 2010).

Oliveira, C.S., Gassol, G., Goula, X., and Susagna, T. (2012). European digital accelerogram data-base: A statistical analysis of engineering parameters of moderate magnitude events. Journal of Seismology, (submitted).

Roca, A., Guéguen, P., Godey, S., Goula, X., Susagna, T., Péquegnat, C., Oliveira, C.S., Clinton, J., Pappaioanou, C. and Zulfikar, C. (2011). The Euro-Mediterranean distributed accelerometric data-base. *In* Earthquake Data in Engineering Seismology. Predictive Models, Data Management and Networks (S. Akkar, P. Gulkan and T. Van

Eck, Editors). Geotechnical, Geological and Earthquake Engineering Book Series. Springer, vol 14, 115-128.

Rueda, J., Mexcua, J. and Garciía Blanco, RM. (2011). Directivity effects of the May 11, 2011 Lorca (Spain) Mw=5.1 earthquake. AGU Fall meating, San Francisco, California, 5-9 December.

Sourieau, A. (2006). Quantifying felt events: A joint analysis of intensities, accelerations and dominant frequencies. Journal of Seismology (2006) **10.**DOI: 10.1007/s10950-006-2843-1.

Tapia, M., Susagna, T., Goula, X. (2007). Curvas predictivas del movimiento del suelo en el oeste del Mediterráneo. 3er Congreso Nacional de Ingeniería Sísmica, 8-11 Mayo 2007, Girona.CD-Rom 17pp.

Trifunac, M.D. y Brady A.G., (1975). A study on the duration of strong earthquake ground motion. BSSA, Vol. 65, 581-626

Uniform Building Code, (1997). International Conference of Building Officials, Wittier, Ca.

Wald, D., Quitoriano, V., Heaton, T. and Kanamori, H. (1999). Relationships beteen peak ground acceleration, peak ground velocity and modified Mercalli intensity in California. Earthquake Spectra, 15, 557-564.